\documentclass[
    ,final            
  ]
  {aipproc}
\usepackage{graphicx}

\layoutstyle{8x11single}

\begin{document}

\title{Strangeness Production at COSY}


\author{Frank Hinterberger}{
  address={HISKP, Universit\"at Bonn, Nussallee 14-16, D 53115 Bonn, Germany}
}

\author{Hartmut Machner}{
  address={Fakult\"at f\"ur Physik, Universit\"at Duisburg-Essen, Lotharstr. 1, D 47048 Duisburg, Germany} 
}

\author{Regina Siudak}{
  address={Inst. Nucl. Physics, PAN, ul. Radzikowskiego 152, 31-342 Krak\'{o}w, Poland}
}

\begin{abstract}
The paper gives an overview of strangeness-production experiments at
the Cooler Synchrotron COSY.
Results on kaon-pair and $\phi$ meson production in
$pp$, $pd$ and $dd$ collisions, 
hyperon-production experiments 
and $\Lambda p$ 
final-state interaction studies 
are presented.
\end{abstract}

\maketitle


\section{Introduction}
The cooler synchrotron COSY \cite{Mai97} 
at the Forschungszentrum J\"ulich in Germany
can accelerate protons and deuterons
up to about 3.7 GeV/c. Both, unpolarized and polarized beams are available.
Excellent beam quality can be achieved using electron- and/or stochastic cooling.
COSY can be used as an accelerator for external target experiments and as storage ring for
internal target experiments.  The strangeness production experiments have been
performed at the internal  spectrometer ANKE by the COSY-ANKE 
collaboration,
at the internal COSY-11 spectrometer by the COSY-11 collaboration, 
at the external TOF facility by the COSY-TOF collaboration 
and at the external BIG KARL spectrometer 
by the COSY-MOMO and COSY-HIRES collaborations.

\section{Kaon-Pair Production Experiments}

\begin{figure}
  \includegraphics[angle=-90,width=0.5\textwidth]{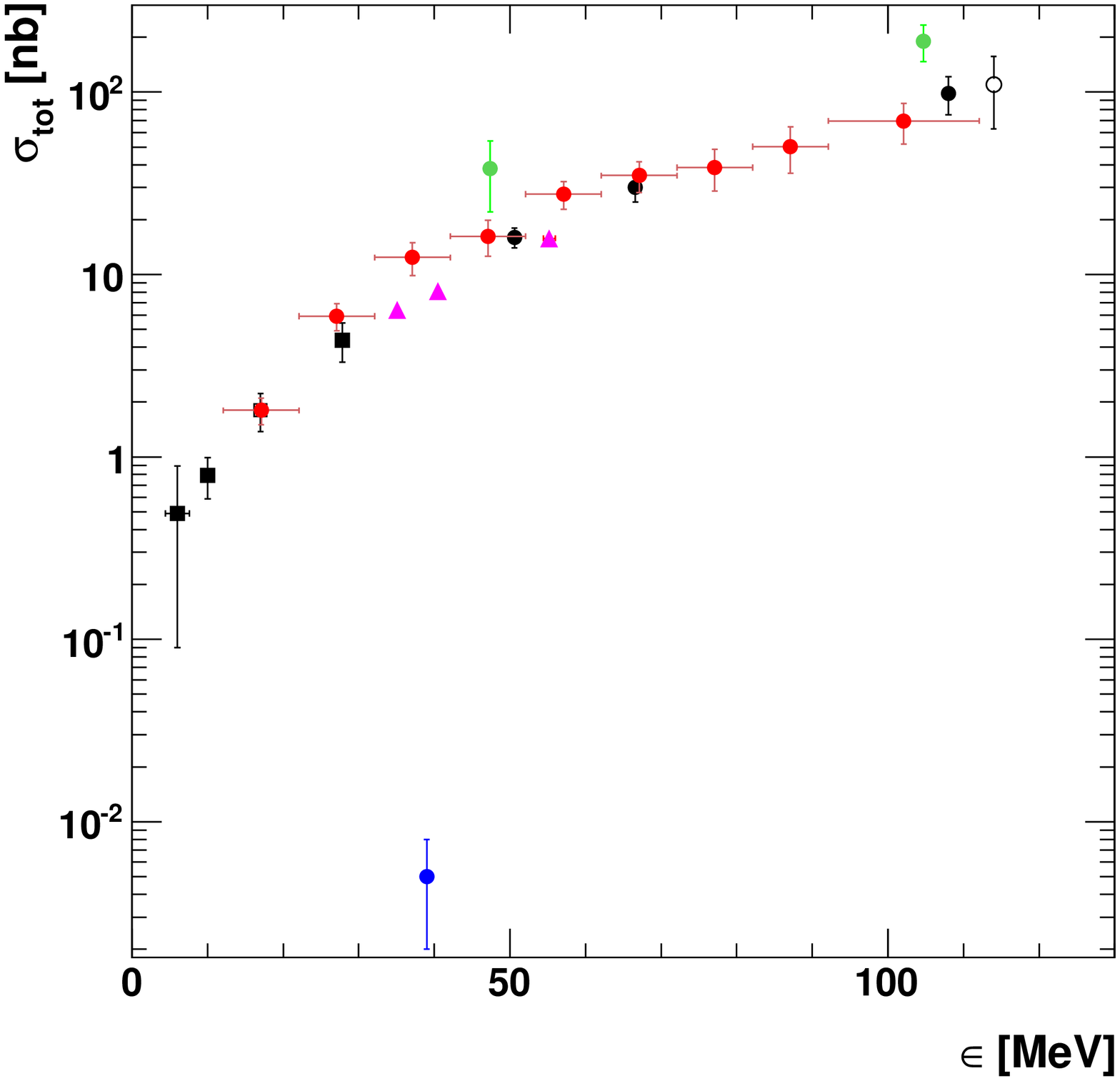}
  \includegraphics[angle=-90,width=0.5\textwidth]{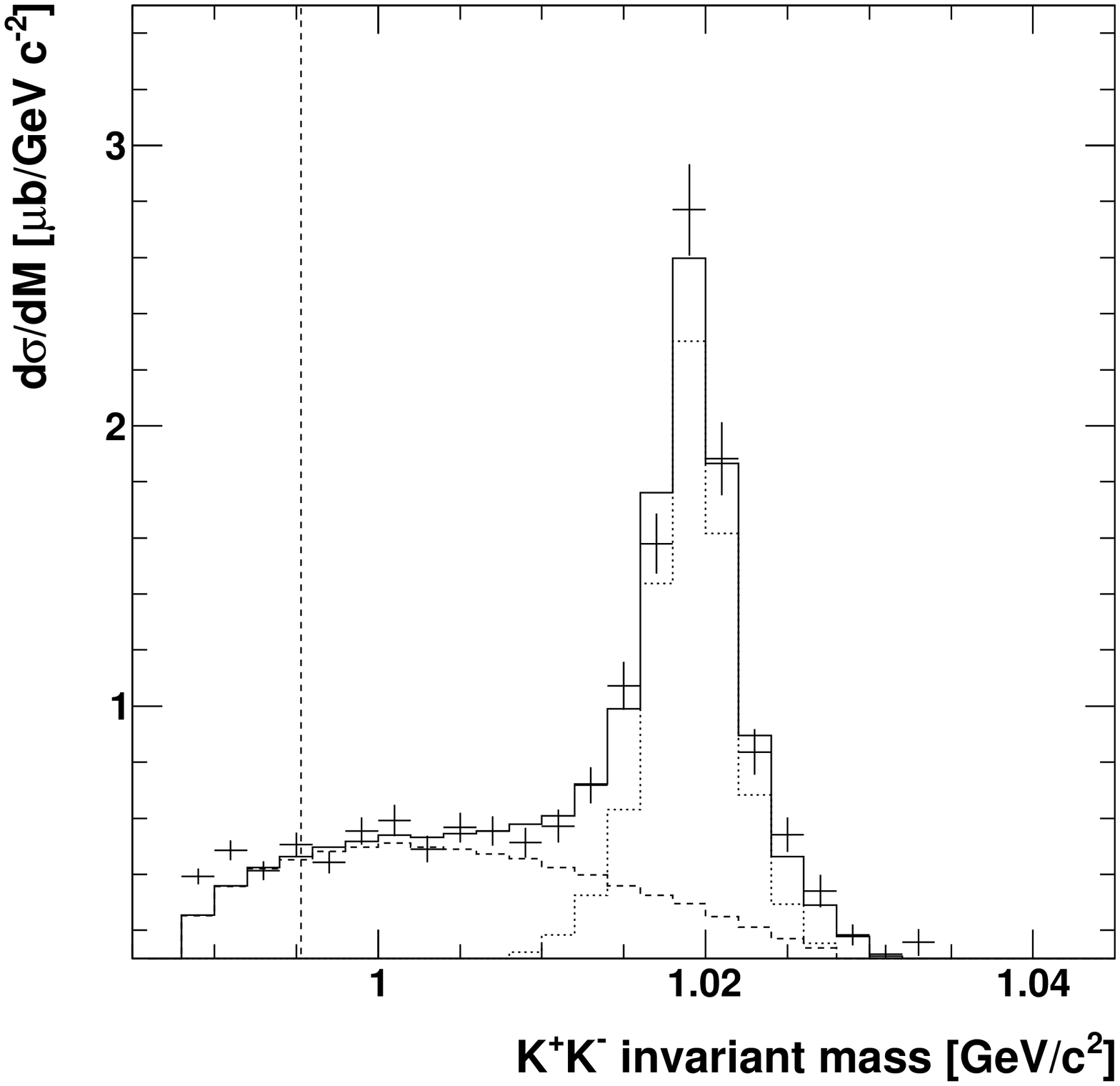}
  \caption{Left: Total cross sections for kaon-pair production as a function of the
excess energy $\epsilon$ measured at COSY. Black: $pp\to ppK^+K^-$ 
\protect{\cite{Wol98,Que01,Win06,Har06,Mae08}}. Green: $pp\to dK^+K^0$
\protect{\cite{Kle03,Dzy06}}. 
Red: $pn\to dK^+K^-$
\protect{\cite{Mae06,Mae09}}. Pink: $pd\to {^3He}K^+K^-$
\protect{\cite{Bel07}}. Blue: $dd\to {^4He} K^+K^-$
\protect{\cite{Yua09}}. 
The high energy $pp\to ppK^+K^-$ result (open circle) has been measured 
at  SATURNE \protect{\cite{Bal01}}.
Right: Invariant mass distribution of $K^+K^-$ pairs measured by COSY-ANKE \protect{\cite{Mae08}} at
$\epsilon=51$~MeV ($T_p=2.65$~GeV). Dotted and dashed histograms represent the $\phi$ and non-$\phi$
contributions. The solid histogram is the sum of both. The $K^0\bar{K}^0$ threshold is indicated by
the dashed vertical  line.
}
  \label{kaon-pair}
\end{figure}

Extensive measurements of kaon-pair production have been performed at several COSY facilities.
Total and differential cross sections are now available for a variety of reactions.
The world data set of
total cross sections for kaon pair production is shown in Fig.~\ref{kaon-pair}, left panel,
as a function of the excess energy $\epsilon$.
The  $pp\to ppK^+K^-$ reaction (black symbols) has been studied by the COSY-11
\cite{Wol98,Que01,Win06} and COSY-ANKE \cite{Har06,Mae08} collaborations
at excitation energies between 3 and 108 MeV.
The closely related reaction $pp\to d K^+ \bar{K}^0$ (green)
has also been studied  by COSY-ANKE \cite{Kle03,Dzy06}.
In addition the reaction $pn\to d K^+ K^-$ (red) has been investigated
by COSY-ANKE \cite{Mae06,Mae09} using a deuterium cluster-jet as neutron target.
The momentum of the non-observed proton spectator and the 
excess energy $\epsilon$ have been reconstructed from
the four-momenta of the detected deuteron and kaon pair.
The $pd\to {^3He} K^+ K^-$ (pink) reaction has been measured by the
COSY-MOMO collaboration \cite{Bel07}.
The $dd\to {^4He} K^+ K^-$ reaction studied by COSY-ANKE is an ideal isospin zero filter.
It could be sensitive to the production of the scalar meson $f_0(980)$.
But the total cross section (blue point) amounts only to
5~pb \cite{Yua09}. The high energy $pp\to ppK^+K^-$ result (open circle) has been measured 
at  SATURNE \cite{Bal01}.

The invariant mass distribution of $K^+K^-$ pairs has been measured by COSY-ANKE \cite{Mae08} at three excess energies
$\epsilon=51, 67$ and 108~MeV. 
The internal ANKE spectrometer detects simultaneously slow positive and negative particles in the side detectors
and fast positive particles in the forward detector.
The spectrum taken at $\epsilon=51$~MeV
($T_p=2.65$~ GeV) is shown in  Fig.~\ref{kaon-pair}, right panel. 
All three spectra show a strong peak
which is due to the production of the $\phi(1020)$ meson which decays with about 50~\% into
$K^+K^-$. The prompt $K^+K^-$ distribution can be
described by a 4-body phase-space distribution modified
by the $pp$ and $K^-p$ final state interaction (FSI).
All three distributions show a low mass enhancement over the fit curves.
This enhancement has been discussed as a coupled channel effect due
to the $K^0\bar{K}^0$ channel  whose threshold (dashed vertical line) is about 8 MeV above
the $K^+K^-$ threshold \cite{Dzy08}.

The  important $K^-p$ FSI has been observed by comparing the invariant mass
distributions of $K^-p$ and $K^+p$ \cite{Mae08,Sil09}. 
The ratio
of $K^-p$ to $K^+p$ production changes by an order of magnitude within 50~MeV.
The same holds true if one compares $K^-pp$ and $K^+pp$.
The $K^-p$ FSI can be described by assuming an imaginary scattering length
of 1.5~fm.

\begin{figure}[h!]
  \includegraphics[angle=-90,width=0.5\textwidth]{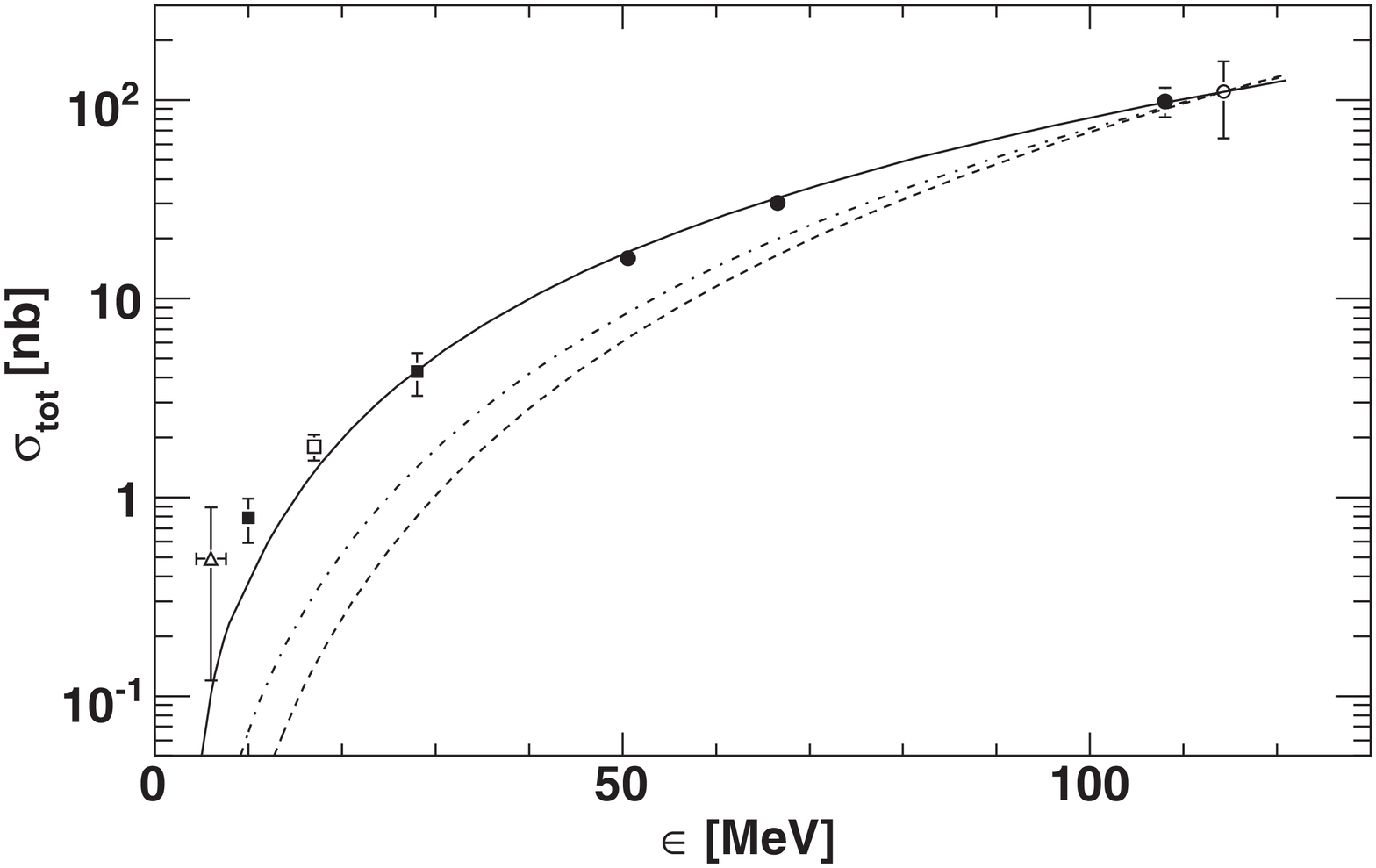}
  \includegraphics[angle=-90,width=0.5\textwidth]{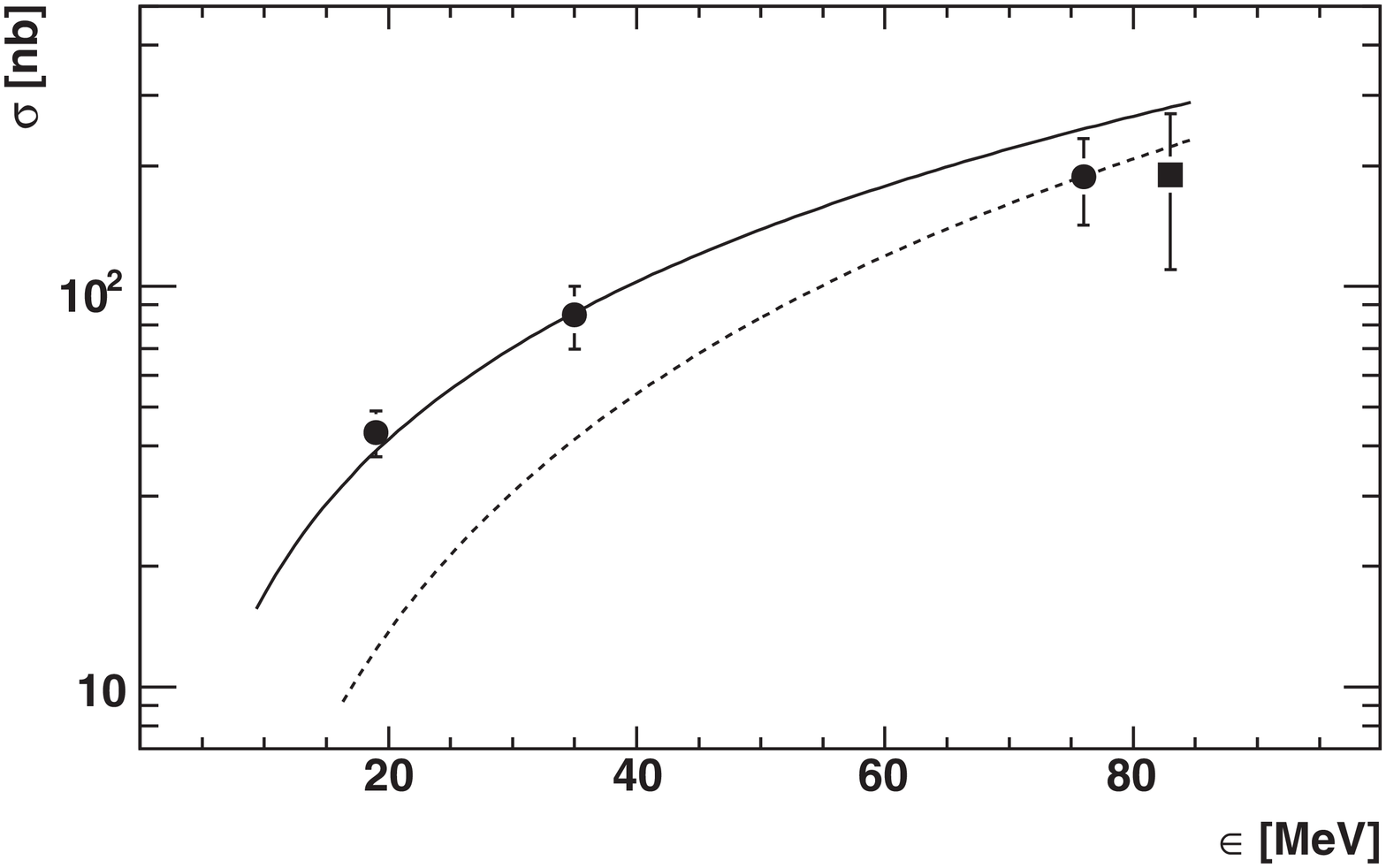}
  \caption{Left: Total cross section for $pp\to ppK^+ K^-$ vs. excess energy $\epsilon$ \protect{\cite{Mae08}}.
Data from COSY-ANKE (closed circles) \protect{\cite{Mae08}}, COSY-11 (open triangle) \protect{\cite{Wol98}},
(open square) \protect{\cite{Que01}}, (closed squares) \protect{\cite{Win06}}
and DISTO (open circle) \protect{\cite{Bal01}}. The dashed curve represents the energy dependence from four-body phase space.
The dot-dashed  includes the $pp$ FSI and the solid curve includes the $K^-p$ FSI.
Right: Total cross sections for $pp\to pp\phi$ vs. excess energy $\epsilon$ \protect{\cite{Har06}}.
Data from COSY-ANKE (closed circles) \protect{\cite{Har06}} and DISTO (closed square) \protect{\cite{Bal01}}.
The dashed curve represents a   three-body phase-space calculation. The solid curve includes the $pp$ FSI.}     
  \label{kaon-pair-phi-tot}
\end{figure}

In Fig.~\ref{kaon-pair-phi-tot}, the left panel shows the total cross sections 
for $pp\to ppK^+K^-$ as a function
of the excess energy $\epsilon$. The data are from COSY-11 \cite{Wol98,Que01,Win06}, 
COSY-ANKE \cite{Har06,Mae08}
and SATURNE-DISTO \cite{Bal01}.
The dashed curve represents a pure 4-body phase-space calculation.
The dot-dashed curve includes the $pp$ FSI and the solid  curve
includes the $K^-p$ FSI. The right panel shows the total
cross section for $\phi$ production as a function of the corresponding excess energy $\epsilon$.
The data cannot be described by a pure three-body phase-space behavior. But the
inclusion of the $pp$ FSI is sufficient to reproduce the energy dependence \cite{Har06}.
Another important result is  the angular distribution of the $\phi$ decay into $K^+K^-$ 
at $\epsilon=18.5$~MeV which is a pure $\sin^2\Theta$
distribution in the Jackson frame. Thus, the $\phi$ meson is tensor polarized with 
$m=\pm 1$ along the beam axis as expected near threshold due to conservation laws \cite{Har06}. 
Comparing the total cross sections for $\phi$- and $\omega$-production
at corresponding  excess energies \cite{Har06,Abd07} yields a ratio $R_{\phi/\omega}$ which is about eight times larger
than the prediction $R_{OZI}=4.2\cdot 10^{-3}$  \cite{Lip76} based on the Okubo-Zweig-Iizuka (OZI) rule.

The $K^+K^-$- and $\phi$-production was also studied by the COSY-MOMO collaboration
at the magnetic spectrograph BIG KARL using the $pd\to {^3He}K^+K^-$ reaction \cite{Bel07}.
The invariant mass distributions measured at three excess energies,
35, 40 and 55 MeV, can be described by pure phase-space distributions. 
Effects due to a possible $K^- {^3He}$ FSI are not visible. In comparison to the 
$pp\to pp \phi$ reaction \cite{Har06} the $\phi$ peak is less pronounced.
The surprising result of the $pd\to {^3He}\phi$ study is that the 
the $\phi$ meson is tensor polarized with 
$m=0$ along the beam axis \cite{Bel07}.
In contrast, the $\omega$ meson is unpolarized in the corresponding reaction
$pd\to {^3He}\omega$ \cite{Sch08}.
The ratio $R_{\phi/\omega}$ of total cross sections for $pd\to {^3He} \phi$ and $pd\to {^3He} \omega$
at corresponding excess energies is about a factor 20  
larger \cite{Bel07} than predicted by the
OZI-rule \cite{Lip76}.

\section{Hyperon-Production Experiments}

Exclusive measurements of
hyperon production have been performed using the associated strangeness reactions
$pp\to K^+ \Lambda p$, $pp\to K^+ \Sigma^0 p$, $pp\to K^+ \Sigma^+ n$ and $pp\to K^0 \Sigma^+ p$
by the COSY-11, COSY-ANKE and COSY-TOF collaborations.
The aim of those measurements is to understand (i) the reaction mechanism,
(ii) to study the effect of $N^*$ resonances and (iii) to study the FSI between
the outgoing nucleon and hyperon.
\begin{figure}[h!]
  \includegraphics[angle=-90,width=\textwidth]{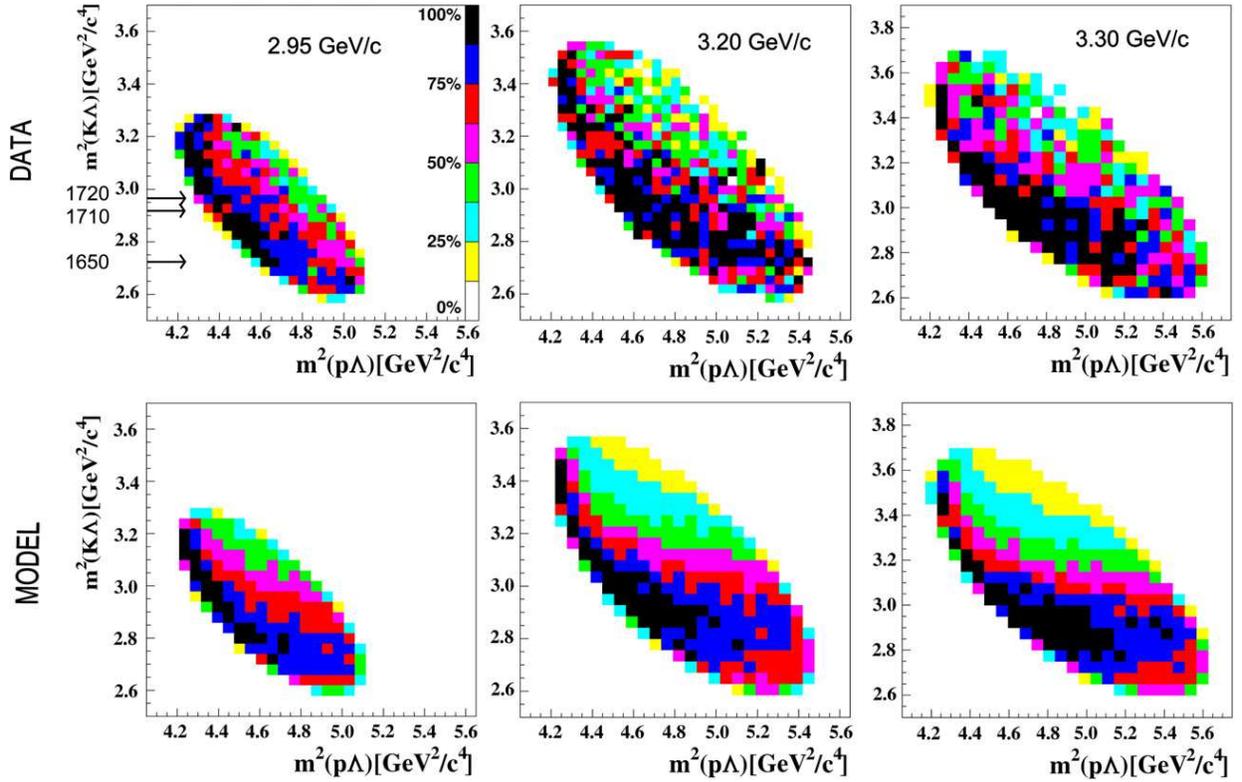}
  \caption{Dalitz plots of the reaction $pp\to K^+ \Lambda p$
measured by COSY-TOF at 2.95, 3.20 and 3.30 GeV/c \protect{\cite{Abd10}}.
The model calculations are performed using the ISOBAR model of Sibirtsev
\protect{\cite{Sib06}}.}
  \label{dalitz-plot}
\end{figure}

\begin{figure}
  \includegraphics[angle=-90,width=0.8\textwidth]{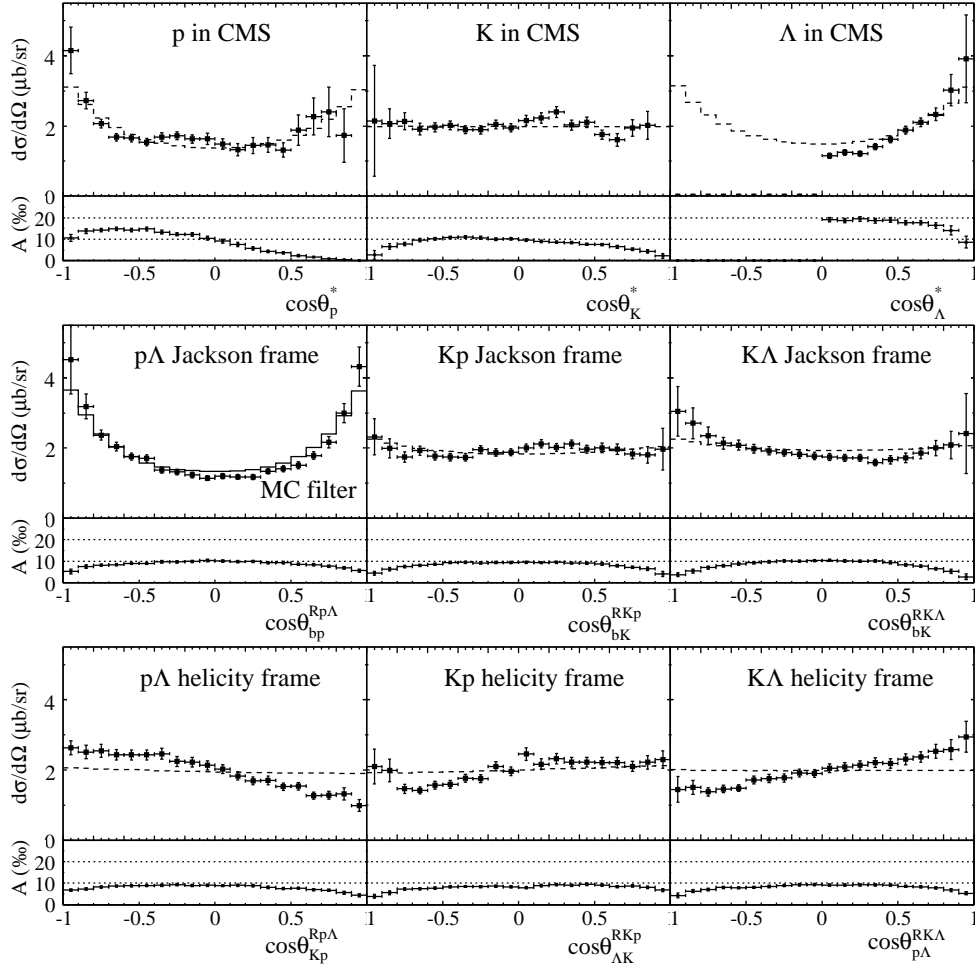}
  \caption{Differential cross sections in the CMS, Jackson and helicity frames
for the reaction $pp\to K^+ \Lambda p$ measured by COSY-TOF at $p_{beam}=3.059$~GeV/c \protect{\cite{Abd11}}. 
Note the differential acceptances $A$(\textperthousand).} 
  \label{ang-distr-lambda}
\end{figure}

The COSY-TOF detector enables exclusive measurements with a large solid angle in the
laboratory system and almost $4\pi$ in the c.m. system.
Recent measurements of the reaction $pp\to K^+ \Lambda p$ at three 
bombarding energies \cite{Abd10} are
shown in Fig.~\ref{dalitz-plot} in the form of Dalitz plots.
The Dalitz plots show strong deviations from a homogeneous phase-space distribution.
The data can be described using  the ISOBAR model of Sibirtsev (concept outlined in \cite{Sib06})
which takes into account the  $\Lambda p$ FSI 
and the contribution of three $N^*$ resonances, $N(1650)$, $N(1710)$ and $N(1720)$.
The exclusive measurements
of $pp\to K^+ \Lambda p$ and $pp\to K^+ \Sigma^0 p$
by COSY-TOF provide also differential cross sections \cite{Abd11}
in the CMS, Jackson and helicity frames, see Fig.~\ref{ang-distr-lambda}.
In addition, first measurements with polarized proton beams have been performed. 
The COSY-TOF data will be used for a detailed partial wave analysis  
by the Bonn-Gatchina PWA group.

\begin{figure}
  \includegraphics[angle=-90,width=0.8\textwidth]{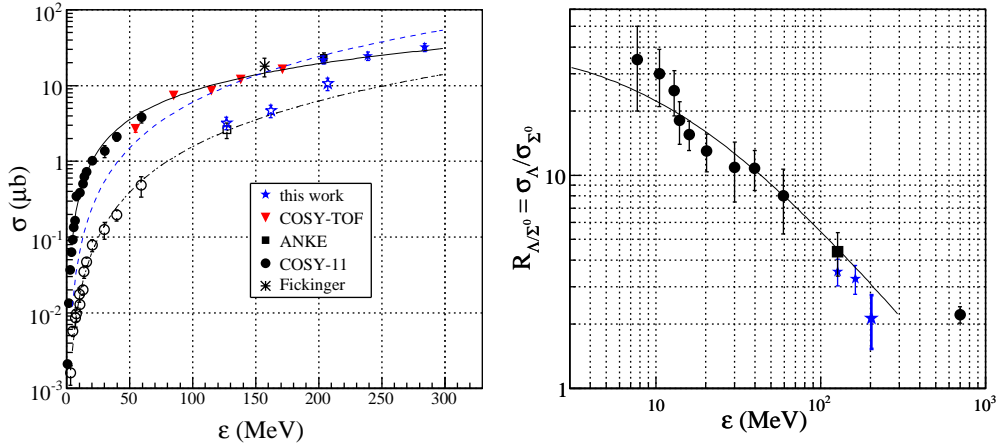}
  \caption{Left: Total cross sections for $pp\to K^+ \Lambda p$ (solid symbols)
 and  $pp\to K^+ \Sigma^0 p$ (open symbols)
\protect{\cite{Abd11,Grz97,Bal97,Bal98,Sev99,Kow04,Abd06,Val07,Fic62}} vs. corresponding excess energy $\epsilon$.
The new data (open and solid crosses, denoted 'this work' in the insert)  are from COSY-TOF \protect{\cite{Abd11}}.
The dashed curves represent a three-body  phase-space $\epsilon^2$ dependence.
The solid curve includes the $\Lambda p$ FSI. 
Right: Ratio of the $pp\to K^+ \Lambda p$ to  $pp\to K^+ \Sigma^0 p$
total cross sections vs. $\epsilon$. The ratio at $\epsilon=700$~MeV is an average 
value calculated from data given in \protect{\cite{Bal88}}. The solid curve takes the $\Lambda p$ FSI into account.}
  \label{canu-schulte-wissermann}
\end{figure}

In Fig.~\ref{canu-schulte-wissermann} 
\cite{Abd11} 
the total cross sections for $pp\to K^+ \Lambda p$ and  $pp\to K^+ \Sigma^0 p$
\cite{Abd11,Grz97,Bal97,Bal98,Sev99,Kow04,Abd06,Val07,Fic62}
are shown as a function of the corresponding excess energy $\epsilon$.
The dashed curves representing a three-body phase-space $\epsilon^2$ dependence  
describe the $pp\to K^+ \Sigma^0 p$ data and
there is no evidence for a $\Sigma^0 p$ FSI.
However, the $pp\to K^+ \Lambda p$ data can only be described by including the $\Lambda p$ FSI.
The right panel shows the ratio of the $pp\to K^+ \Lambda p$ to  $pp\to K^+ \Sigma^0 p$
total cross sections
as a function of the excess energy. The new experimental values from COSY-TOF \cite{Abd11}
confirm the general trend towards a ratio of 2.2 measured at high energies \cite{Bal88}.
The large increase of the ratio towards low energies  is well described by the $\Lambda p$ FSI.

The first measurement  of the reaction $pp\to K^+ \Sigma^+ n$ close to threshold
has been performed by COSY-11 \cite{Roz06}, by detecting the $K^+$  in coincidence
with the neutron. The total cross sections at $\epsilon=13$ and 60 MeV were astonishingly high
compared to those for $pp\to K^+ \Sigma^0 p$.
Therefore, the  energy dependence of the reaction $pp\to K^+ \Sigma^+ n$ 
close to threshold has been studied recently by COSY-ANKE
\cite{Val07,Val10} by
detecting $K^+$ in coincidence with $\pi^+$ from the decay of the
$\Sigma^+$ and by detecting $K^+ p$ coincidences.
In contrast to the COSY-11 results \cite{Roz06} the total cross sections for $pp\to K^+ \Sigma^+ n$
are very much smaller (factor 50 at $\epsilon=60$~MeV) 
and
slightly smaller than those for  $pp\to K^+ \Sigma^0 p$ ($R(\Sigma^+/\Sigma^0)=0.7\pm0.1$).
The energy dependence can be described by a three-body phase-space calculation
\cite{Val10}, 
with no evidence for a  $\Sigma^+n$ FSI.
Details can be found in the contribution of Yu. Valdau to these proceedings.

A recent determination of the total cross section for $pp\to K^+ \Sigma^+ n$
at $\epsilon=102.6$ MeV from inclusive $K^+$ data by COSY-HIRES \cite{Bud11}
yielded also a  value 
smaller than expected from the COSY-11 results. But the result is
in conflict with the exclusive measurement by COSY-ANKE \cite{Val07,Val10}.
Possible reasons for this discrepancy 
are discussed in \cite{Val11}. 
\begin{figure}[h!]
  \includegraphics[width=0.5\textwidth]{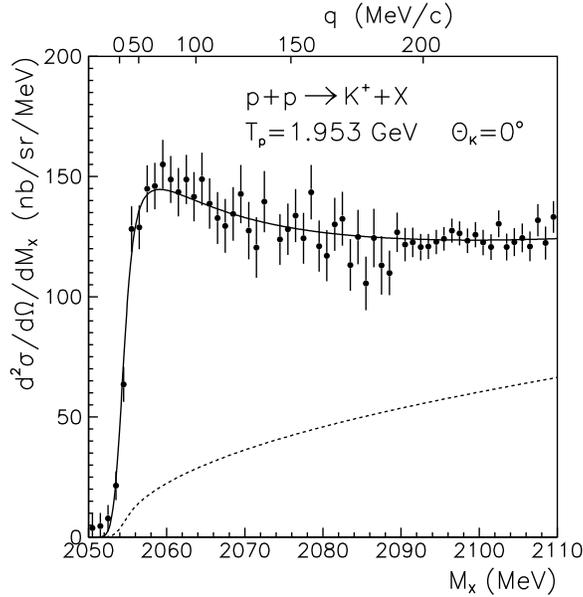}
  \caption{Missing mass spectrum of the reaction $pp\to K^+ \Lambda p$
measured at $T_p=1.953$~GeV and $\Theta_K=0^{\circ}$ \protect{\cite{Bud10}}.
Solid curve: fit with spin-averaged
effective range parameters. Dashed curve: phase-space distribution.}
  \label{hires-spectrum}
\end{figure}

The $\Lambda p$ FSI in the reaction $pp\to K^+ \Lambda p$ has been studied with a high invariant mass resolution
by the COSY-HIRES collaboration \cite{Bud10}. The kaons were detected 
at $0^{\circ}$
using the magnetic spectrograph BIG KARL \cite{Dro98}. The double differential cross section is shown in Fig.~\ref{hires-spectrum}
as a function of the missing mass, i.e. the invariant mass of the $\Lambda p$ system.
The spectrum is characterized by a huge FSI enhancement near the $\Lambda p$ threshold.
A narrow $S=-1$ dibaryon resonance predicted near 2100~MeV \cite{Aer85}
is not visible.
The data can be described by factorizing the reaction amplitude in terms of a production amplitude
and a FSI enhancement factor (solid curve in Fig.~\ref{hires-spectrum}). 
Taking for the enhancement the inverse Jost function
yields the spin-averaged effective range parameters of the $\Lambda p$ interaction,
i.e. the scattering length $\bar{a}=-2.43^{+0.16}_{-0.17}$~fm and the effective range $\bar{r}=2.21\pm 0.16$~fm.
The dashed curve is the corresponding phase-space distribution without FSI enhancement.
Taking the free $\Lambda p$ scattering data \cite{Ale68,Sec68} in a combined fit into account
yields the possibility to disentangle the spin singlet and triplet effective range parameters
of the $\Lambda p$ system \cite{Bud10}. However, 
this result should be considered with respect to the theoretical uncertainties of the 
Jost-function approach \cite{Gas05}.
A direct determination of the spin singlet and triplet effective range parameters requires
polarization measurements as proposed by Gasparyan et al. \cite{Gas04} and planned by COSY-ANKE and COSY-TOF.


\begin{theacknowledgments}
The results presented here are due to the efforts of the
collaborations COSY-ANKE, COSY-11, COSY-HIRES, COSY-MOMO and COSY-TOF
and the COSY machine crew. 
The authors thank K.-Th. Brinkmann, H. Clement, W. Eyrich, H. Freiesleben, M. Hartmann, J. Ritman, E. Roderburg, M. Schulte-Wissermann,
A. Sibirtsev, H. Str\"oher and Yu. Valdau
for their help and clarifying discussions. 
\end{theacknowledgments}



\bibliographystyle{aipproc}   
\end{document}